\documentclass[aps,reprint,onecolumn, showpacs,notitlepage]{revtex4-1}
\usepackage{epsfig,graphics,amssymb,amsmath,subeqnarray,color,bm}
\usepackage[toc,page]{appendix}
\usepackage{mathtools}

\usepackage{xcolor}
\usepackage{float}
\usepackage{cancel}
\usepackage{amsmath}

\numberwithin{equation}{section}
\usepackage{graphicx}
\definecolor{navyblue}{RGB}{0,0,128}
\definecolor{dodgerblue}{RGB}{30,144,255}
\definecolor{darkgrey}{RGB}{169,169,169}
\definecolor{deepskyblue}{RGB}{0, 191, 255}

 \def\boldsymbol{\bm}

\def \d{\,\text{d}}

\def \bgammad{\dot{\boldsymbol{\gamma}}}

\def \bnabla{\boldsymbol{\nabla}}

\def \btau{\boldsymbol{\tau}}

\def \bU{\mathbf{U}}

\def \be{\mathbf{e}}

\def \br{\mathbf{r}}

\def \bu{\mathbf{u}}

\begin{document}
\title{A note on higher order perturbative corrections to squirming speed in weakly viscoelastic fluids}
\author{Charu Datt} \email{Electronic mail: charudatt@alumni.ubc.ca}
\affiliation{Department of Mechanical Engineering, 
University of British Columbia,
Vancouver, BC, V6T 1Z4, Canada
}

\author{Gwynn J. Elfring}
\affiliation{Department of Mechanical Engineering, 
University of British Columbia,
Vancouver, BC, V6T 1Z4, Canada
}
\date{\today}

\begin{abstract}
Many microorganisms swim in fluids with complex rheological properties. Although much is now understood about motion of these swimmers in Newtonian fluids, the understanding is still developing in non-Newtonian fluids\textemdash this understanding is crucial for various biomimetic and biomedical applications.  Here we study a common model for microswimmers, the squirmer model, in two common viscoelastic fluid models, the Giesekus fluid model and fluids of differential type (grade three),  at zero Reynolds number. Through this article we address a recent commentary that discussed suitable values of parameters in these model and pointed at higher order viscoelastic effects on the squirming motion.  
\end{abstract}

\maketitle

\section{Introduction}
With ideas of non-invasive surgery, targeted drug delivery, and other biomimetic applications \citep{nano, gao2014synthetic}, the understanding of motion of microswimmers in complex fluids has become imperative. Subsequently,  many recent articles have focussed on motion of microswimmers in complex fluids (see reviews \citep{sznitman15, elfring2015theory}). While biological fluids may demonstrate many non-Newtonian fluid properties, one common property is viscoelasticity \citep{viscoelastic_blood, Lai_mucus}. We consider this property in this article. 

Viscoelastic fluids show both viscous and elastic properties, and retain memory of their flow history \citep{bird1987dynamics}. Recent experimental studies on biological swimmers  \citep{arratia_prl, Arratia_science, Arratia_chlamydomonas} have addressed how the organism may change its swimming stroke as it ``senses" the viscoelasticity of the fluid media. Elastic stresses in the fluid may also directly contribute to changes in the swimming speed of the swimmer for constant swimming strokes.  In this work, we concern ourselves with the theoretical models that may be used to understand the swimmer and its motion in viscoelastic fluids. One model of microswimmers conducive to theoretical treatment is the squirmer model \citep{lighthill1951}. The model, developed by \citet{lighthill1951} and \citet{Blake1971},  consists of a rigid body that generates thrust due to the presence of (apparent) slip velocities on its surface. It has been used to understand various single and collective behaviours of microswimmers in Newtonian fluids \citep{Pedley_review}. In viscoelastic fluids, \citet{Lailai} studied the motion of squirmers using numerical simulations and found that all squirmers---pushers, pullers and neutral swimmers---swim slower than in a Newtonian fluid for a wide range of values of the Weissenberg number (measure of viscoelasticity in the fluid). Later, \citet{Maffettone} using a theoretical approach showed that in fact for very small values of the Deborah (Weissenberg) number not considered in the work of \citet{Lailai} pusher swimmers swim faster, puller swimmers slower and neutral swimmers at the same speed as in a Newtonian fluid.  We note that in these studies, as will be the case in the present study, the swimming speed is compared for the same swimming stroke in viscoelastic and Newtonian fluids. 

The work of \citet{Maffettone} used a second order fluid model to study weakly viscoelastic effects on squirming motion. The use of the second order fluid model with parametric values as chosen by \citet{Maffettone} was critiqued by \citet{christov16} who argued that the parametric values be chosen in accordance with thermodynamic constraints and recommended the use of other viscoelastic models which ``better elucidate the transient effects of fluid viscoelasticity on a squirmer".   \citet{decorato16b} then showed that in fact using the Giesekus model to study weakly viscoelastic effects, to $\mathcal{O} \left( De \right) $, gives results identical to those previously obtained by them using the the second order fluid model. The motivation for this work in large part is due to this discussion; here we study the squirming motion to higher orders in Deborah number both in the Giesekus fluid and in fluids of differential type. We find that unlike in a second-order fluid that obeys thermodynamic constraints, weak viscoelastic contributions to the squirming speed are non-zero in a fluid of grade three (third-order fluid) obeying thermodynamic constraints. These contributions are qualitatively different to those obtained due to viscoelasticity as modelled by the Giesekus fluid. 

In the following, we briefly discuss the squirmer model and the second order fluid model with the points of contention, and then present our results.

 \section{Theoretical framework}
\subsection{The squirmer model}
The spherical squirmer model consists of a sphere with prescribed axisymmteric surface velocities (surface velocities may be thought of as originating from surface distortions in biological microswimmers like \textit{Opalina}) which generate thrust forces to propel the swimmer \citep{lighthill1951, Blake1971}. We consider only a tangential surface velocity on the swimmer (the swimmer maintains its shape) so that the surface velocity  $\bu^S = u^S_{\theta} \be_{\theta}$, where $u^S_{\theta}$ can be expressed as

\begin{equation}
u^S_{\theta} = \Sigma_{l = 1}^{\infty} B_l V_l \left( \theta \right),
\end{equation} 
using $V_l \left( \theta\right) = -\left(2/ (l (l+1)) \right) P^1_l \left( \cos{\theta}\right)$; $ P^1_l \left( \cos{\theta}\right)$ are associated Legendre polynomials of the first kind, and $\theta$ is the polar angle measured from the axis of symmetry \citep{Blake1971}. The coefficients $B_l$ are generally referred to as squirming modes. In Newtonian fluids, the swimming speed of the squirmer is due to just the first mode, $U_N = 2/3 B_1$, and the second mode $B_2$ gives the stresslet due to the squirmer \citep{ishikawa}. As velocities due to the higher modes decay faster than the first two modes (in fact $B_2$ gives the slowest decaying spatial contribution to the flow field), and since higher modes do not contribute to the swimming speed, in Newtonian fluids, often only the first two modes are considered, i.e., $B_n = 0$ for $n> 3$. For the purpose of this study, in accordance with the bulk of literature in the field \citep{Pedley_review}, we too consider only the first two modes. At this point we feel it is important to note that in general considering only the first two modes in complex fluids may be problematic as shown in the recent works by \citet{jfmCharu, datt17}. To the interested reader we also point to the description of non-axisymmetric squirming modes in Newtonian fluids by \citet{Pak2014}. 

When the ratio $\beta = B_2/ B_1$ is negative, the squirmer generates thrust from its rear end, like the bacterium \textit{E. coli.}; when $\beta> 0$ the thrust is generated from the front end, as in the algae \textit{Chlamydomonas}. When $\beta = 0$, the thrust and drag centres coincide. The three types of squirmers are called pushers, pullers, and neutral swimmers \citep{Pedley_review}. 

\subsection{The second order fluid model}
The deviatoric stress in a second order fluid  \citep{rajagopal} is given by
\begin{equation}
\btau = \eta \mathbf{\mathsf{A}}_1 + \alpha_1 \mathbf{\mathsf{A}}_2 + \alpha_2 \mathbf{\mathsf{A}}_1^2, 
\label{second-order}
\end{equation}
where
\begin{equation}
\begin{aligned}
\mathbf{\mathsf{A}}_1 &\equiv \mathbf{\mathsf{L}} + \mathbf{\mathsf{L}}^{T},\\
\mathbf{\mathsf{A}}_n &\equiv \frac{\text{D} \mathbf{\mathsf{A}}_{n-1}} {\text{D}t} + \mathbf{\mathsf{L}}^{T} \mathbf{\mathsf{A}}_{n-1} + \mathbf{\mathsf{A}}_{n-1} \mathbf{\mathsf{L}},  
\end{aligned}
\end{equation}
with $\mathbf{\mathsf{L}} ^T = \bnabla \bu$ \citep{tanner2000engineering, fosdick}. Here $\eta$ is the shear viscosity and $\alpha_1$ and $\alpha_2$ are material moduli. The second order fluid model has been used to study the first effects of viscoelasticity on the motion of both passive and active particles (see for e.g., \citep{lailai_snowman, two_sphere_viscoelastic, brunn_review}). However, there has been much discussion on the permissible values of $\alpha_1$ and $\alpha_2$ in the model. \citet{Dunn1974} have shown that considering \eqref{second-order} as exact, the model is consistent with thermodynamics when 

\begin{align}
&\eta \geq 0, \\
&\alpha_1 \geq 0,\\ 
&\alpha_1 + \alpha_2 = 0. 
\end{align}

However, often these constraints, citing experimental investigations (incorrectly according to \citep{rajagopal}), are not strictly adhered to. In particular,  $\alpha_1$, which corresponds to the first normal stress difference coefficient, is generally taken negative \citep{rajagopal}.

\subsection{The reciprocal theorem}

The reciprocal theorem of low Reynolds number hydrodynamics \citep{happel2012low} can be used to calculate the first effects of the fluid rheology on the swimming speed of microswimmers \citep{Lauga_theorem}. 
The details of the reciprocal theorem for the specific case of squirmers in viscoelastic fluids may be found among others in the works of \citet{High_Deborah}, \citet{Maffettone} and \citet{datt17}. 

Consider the weakly non-linear fluid of the form \citep{Lauga_theorem}
\begin{equation}
\btau = \eta \bgammad + \varepsilon \boldsymbol{\Sigma} \left[\bu \right],
\label{weak_fluid}
\end{equation}
where $\btau$ is the deviatoric stress, $\eta$ is the shear viscosity, and $\bgammad$ is the strain rate tensor so that the first term on the right hand side in \eqref{weak_fluid} gives the Newtonian contribution. Here $\varepsilon$ is the small parameter that quantifies the deviation from the Newtonian behaviour and $\boldsymbol{\Sigma}$ gives the non-Newtonian contribution. Using the reciprocal theorem in this fluid, one obtains the translational velocity of a squirmer with radius $a$ as \citep{datt17}

\begin{equation}
\bU = -\frac{1}{4 \pi a^2} \int_S \bu^S \d S - \varepsilon \frac{1}{8 \pi \eta } \int_{\mathcal{S}} \boldsymbol{\Sigma} : \left( 1+ \frac{a^2}{6} \nabla^2 \right) \bnabla \mathbf{\mathsf{G}} \d V,
\label{swim_vel}
\end{equation}
where $\mathbf{\mathsf{G}} = \left(1/ r \right)\left( \mathbf{\mathsf{I}} + \br\br/ r^2 \right) $ is the Oseen tensor \citep{datt17}. 

\section{Results and discussion}

\citet{Maffettone} studied the motion of a squirmer in a second-order fluid. Considering only small deviations from the Newtonian behaviour, they expanded all flow quantities in the small parameter $\varepsilon = De$, where Deborah number $De = -\alpha_1 B_1 / \eta a$ is a measure of the relaxation time scale of the fluid to the characteristic time scale of the flow (note that for steady surface slip velocity squirmers, the Deborah and Weissenberg numbers are equivalent \citep{Poole_De}).  Note that \citet{Maffettone} assumed $\alpha_1 < 0$, in contradiction with the thermodynamic stability criterion as pointed out by \citet{christov16}. The thermodynamic constraint $\alpha_1 + \alpha_2 = 0$ was also relaxed. \citet{Maffettone} found that the perturbation calculations predicted that pushers swim faster, pullers slower and neutral swimmers at the same speed as in Newtonian fluids, provided that the swimming gait remains unchanged between the viscoelastic and Newtonian fluids. Their numerical simulations in a Giesekus fluid found the analytical results to hold up to $De \approx 0.02$ \citep{Maffettone}.  It was commented that the deviation of the analytical results from those from numerical simulations at larger $De$ was due to higher order viscoelastic effects that were neglected in the analytical results where only $\mathcal{O} \left( De\right)$ corrections were analysed \citep{Maffettone}. 

The critique of the work of \citet{Maffettone} by \citet{christov16} was centred about the former not respecting the thermodynamic constraints of the second-order fluid model. In particular, \citet{christov16}  remarked that since $\alpha_1 + \alpha_2$ should be equal to zero, most corrections to flow quantities (but pressure) including the swimming speed of the squirmer will be zero, since all these corrections are proportional to the sum $\alpha_1 +  \alpha_2$.  Citing \citep{Dunn1974}, \citet{christov16} also pointed out that for $\alpha_1< 0$ a steady solution to the problem should not be expected. Finally, \citet{christov16} suggested calculating corrections to the swimming motion with the thermodynamic constraints (meaning going to higher powers in $De$ for any non-zero contributions) or  using a different viscoelastic model, such as the upper-convected Maxwell model. 

\citet{decorato16b} showed that even with using a more involved model like the Giesekus fluid model (which reduces to the upper-convected Maxwell model for a choice of a model parameter), one obtains equations identical to the second-order fluid in the limit of small $De$ at $\mathcal{O} \left( De\right)$. Further, for its permissible values, the Giesekus fluid gives identical results to those from the second order fluid as used by \citet{Maffettone}.  In fact, they maintain that the second order fluid model should be seen as an approximate to more complex viscoelastic models in slow and nearly steady flows (and therefore \eqref{second-order} not be seen as exact).  Perhaps, in order to avoid any confusion, one may restrict the use of the term ``second-order fluid model''  only when it is treated as an exact model obeying the thermodynamics constraints; where a slow and nearly steady flow approximation is used one can start with a more involved model and reduce it to simpler constitutive equations at each order in the perturbation series in $De$. Below we use this terminology and study the squirmer in a Giesekus fluid and in fluids of grade $n$ (second order fluid is a fluid of grade two) and calculate the corrections to the swimming speed in these fluids up to higher orders in $De$.

\subsection{Giesekus fluid}

The polymeric stress in a  Giesekus fluid is given as \citep{morozov2015introduction}
 
\begin{equation}
\btau_p + \lambda \buildrel \nabla \over {\btau_p} + \alpha_m \frac{\lambda}{\eta_p} \btau_p \cdot \btau_p = \eta_p \bgammad, 
\label{eq_giese}
\end{equation}
where the mobility factor $\alpha_m$ must take values between 0 and 1/2 \citep{morozov2015introduction, Lailai}. The total deviatoric stress in the fluid is $\btau = \btau_s + \btau_p$ where $\btau_s = \eta_s \bgammad$ is the contribution from the Newtonian solvent. The total viscosity in the fluid $\eta = \eta_s + \eta_p$, and here we consider the case when $\zeta = \eta_s / \eta = 0$. Note that when $\zeta = 0$ and $\alpha_m = 0$, \eqref{eq_giese} reduces to the upper-convected Maxwell fluid model \citep{morozov2015introduction}. 

\begin{figure}
\includegraphics[width = 0.5\textwidth]{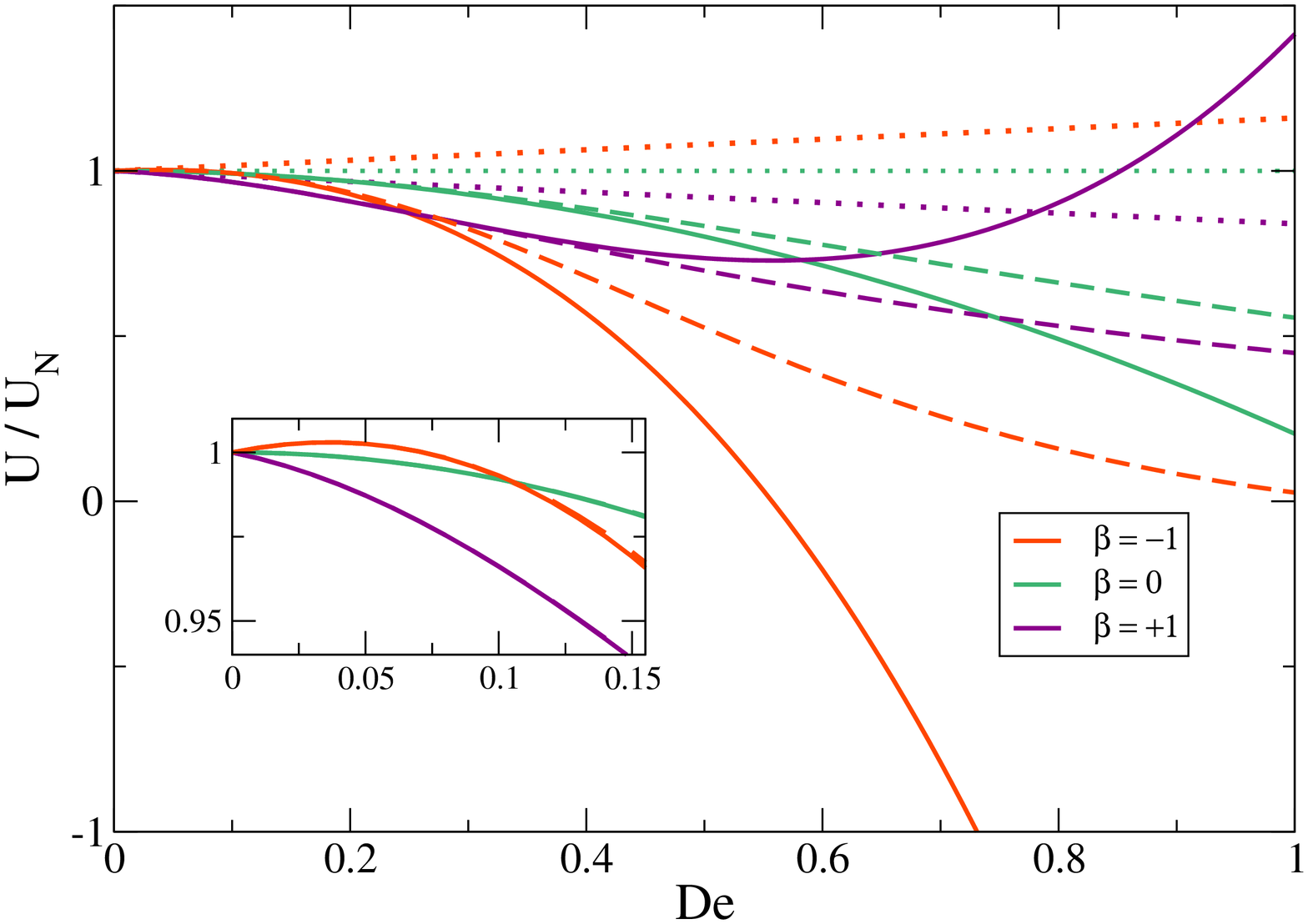}
\caption{The solid lines include corrections up to $\mathcal{O} \left( De^3\right)$. The dashed lines are Pad\'{e} approximations to the series for the speeds in the text.  The dotted lines include only $\mathcal{O} \left( De\right)$ corrections. The addition of the higher order modes decreases the speeds of the squirmers. As seen here, all squirmers at large values of $De$ swim slower than in a Newtonian fluid, as found in the numerical work of \citet{Lailai}. }
\label{squirmer_deborah_fig}
\end{figure}

We non-dimensionalise equations by scaling lengths by the squirmer radius $a$, velocities with the first squirming mode $B_1$, and stresses with $\eta B_1 / a$, and obtain the dimensionless constitutive equation 
\begin{equation}
\btau^* + De \buildrel \nabla \over {\btau^*} + \alpha_m De \btau^*\cdot \btau^* = \bgammad^*,
\end{equation}
where the Deborah number $De = \lambda B_1/ a$. Henceforth, we drop the stars for convenience. We expand all flow quantities in a regular perturbation expansion in $De$, and using standard methods to calculate the flow fields in Stokes flow \citep{happel2012low} obtain the swimming speed of the squirmer, up to $\mathcal{O} \left( De^3\right)$, 
\begingroup\makeatletter\def\f@size{9.6}\check@mathfonts
\def\maketag@@@#1{\hbox{\m@th\large\normalfont#1}}%
\begin{equation}
\begin{aligned}
U &= \frac{2}{3} + \frac{2}{15} \beta \left( -1 + \alpha_m \right) De  \\&+ \frac{  \beta^2 \left( -20568 - 98136 \alpha_m + 65266 \alpha_m^2 \right) + 84 \left( -193 + 176 \alpha_m \left( -3 + 2 \alpha_m\right)\right)}{45045} De^2 \\ 
&+ \frac{\beta} {482431950} \bigg( {170 \left(3005646 + \alpha_m \left( 6190100 + 3 \alpha_m \left( -10014053 + 4815243 \alpha_m \right) \right) \right)}  \\ &+ {\beta^2  \big( 224764987 + \alpha_m \left( 1298121442 + 3 \alpha_m \left( -1659132865 + 875 113 652 \alpha_m \right) \right)\big) } \bigg)De^3.
\label{asymp_eq}
\end{aligned}
\end{equation} \endgroup
At this point putting values of $\beta$ becomes instructive;  we choose $ \beta = -1$ for pushers, $0$ for neutral squirmers, and $1$ for puller type squirmers and $\alpha_m = 0.2$. These values correspond to the values used in the work of  \citet{Maffettone}. From \eqref{asymp_eq} we find, for pushers, 
\begin{equation}
\frac{U}{U_N} = 1 + 0.16 De - 2.05 De^2 - 2.62 De^3, 
\label{eq_pushers}
\end{equation}
for pullers, 
\begin{equation}
\frac{U}{U_N} = 1 - 0.16 De - 2.05 De^2 + 2.62 De^3,
\label{eq_pullers}
\end{equation}
and for neutral squirmers, 
\begin{equation}
\frac{U}{U_N} = 1 - 0.80 De^2. 
\label{eq_neutral}
\end{equation}
The swimming speeds in \eqref{eq_pushers}, \eqref{eq_pullers}, and \eqref{eq_neutral} are plotted in figure \ref{squirmer_deborah_fig} with respective Pad\'{e} approximant $P^1_2 \left(De\right)$ \citep{bender2013advanced}.   When corrections up to only $\mathcal{O} \left(De\right)$ are considered, we note that pushers swimmer swim faster, pullers slower and neutral swimmers at the same speed as in a Newtonian fluid; this is shown in the work of \citet{Maffettone}. With terms up to $\mathcal{O} \left( De^3\right)$, we note that all the squirmers swim slower (except for very small values of $De$) than in a Newtonian fluid as found in the numerical work of \citet{Lailai}. Clearly, the inclusion of higher order terms changes the theoretical predictions significantly. 

One may calculate the higher order terms in the expansion to predict results for larger values of $De$. This is done by  \citet{housiadas2016}, up to $\mathcal{O} \left( De^8 \right)$, for steady sedimentation of a passive sphere in a viscoelastic fluid. They also quantify when the results from the series should not be considered (using positive definiteness of the conformation tensor). \citet{Taylor_improvement} and \citet{elfring2015theory} also performed a higher order perturbation analysis, using techniques to improve the convergence properties of the series, for the swimming speed of a two-dimensional swimming sheet where the small parameter was the amplitude of the waves on the sheet. We have not pursued these endeavours here, for the motivation for this study was to see the differences between the different viscoelastic models considering only the first few terms.

The results in the foregoing use the Giesekus model for viscoelasticity. They would remain qualitatively the same if one were to use the upper-convected Maxwell model.   But what happens to a squirmer in a fluid of grade $n$, when the fluid is ``regarded as a fluid in its own right, not necessarily an approximation to any other one" \citep{truesdell} ?

\subsection{A fluid of grade three}
Consider a fluid of grade three \citep{fosdick}:  
\begingroup\makeatletter\def\f@size{10.2}\check@mathfonts
\def\maketag@@@#1{\hbox{\m@th\large\normalfont#1}}%
\begin{equation}
\btau = \eta \mathbf{\mathsf{A}}_1 + \alpha_1 \mathbf{\mathsf{A}}_2 + \alpha_2 \mathbf{\mathsf{A}}_1^2 + \beta_1 \mathbf{\mathsf{A}}_3 + \beta_2 \left[ \mathbf{\mathsf{A}}_1 \mathbf{\mathsf{A}}_2 + \mathbf{\mathsf{A}}_2 \mathbf{\mathsf{A}}_1\right] + \beta_3 \left( tr \mathbf{\mathsf{A}}_1^2 \right)\mathbf{\mathsf{A}}_1,
\label{eq_third_order_fluid}
\end{equation} \endgroup
where $\mu$, $\alpha_1$, $\alpha_2$, $\beta_1$, $\beta_2$, and $\beta_3$ are material moduli. 
The equation is dimensional. 
Thermodynamics stipulates \citep{fosdick} that 
\begin{equation}
\begin{aligned}
&\eta \geq 0 \quad \alpha_1\geq 0, \quad \vert \alpha_1 + \alpha_2 \vert \leq \sqrt{24 \mu \beta_3}, \\
&\beta_1 = 0 \quad \beta_2 = 0 \quad \beta_3 \geq 0. 
\label{thermo_constraints}
\end{aligned}
\end{equation}

We scale flow quantities as before, and consequently, equation \ref{eq_third_order_fluid} with \ref{thermo_constraints}, in its dimensionless form, becomes  

\begin{equation}
\btau = \bgammad + De \left[ \buildrel \Delta \over \bgammad + \mathcal{Q} \bgammad \cdot \bgammad \right] + De^2 \left[\text{tr} \left(\bgammad \cdot \bgammad\right)  \mathcal{P} \bgammad \right] ,
\end{equation}
where $De = \alpha_1 B_1 / \mu a $, $\mathcal{Q} = \alpha_2 / \alpha_1$ and $\mathcal{P} = \beta_3 \mu / \alpha_1^2$. $ \buildrel \Delta \over \bgammad $ is the lower convected derivative of $\bgammad$ \cite{morozov2015introduction}, denoted by $\mathbf{\mathsf{A}}_2$ in equation \ref{eq_third_order_fluid}.  We expand all flow quantities in a regular perturbation expansion of $De$ and calculate the propulsion speed (dimensionless), which comes to be, to $\mathcal{O} \left( De^2\right)$, 

\begin{equation}
\begin{split}
U = \frac{2}{3} -& \frac{2}{15} \beta \left( 1+ \mathcal{Q}\right) De \\& - \frac{2 \beta^2 \left( 1+ \mathcal{Q}\right) \left( 161+ 559\mathcal{Q} \right) - 48 \left( 616+ 1383 \beta^2\right) \mathcal{P} }{45045} De^2. 
\end{split}
\label{third_swimming}
\end{equation}
Note that when $\mathcal{P} = 0$, we obtain a fluid of grade two, where $1+ \mathcal{Q} = 0$ (\ref{thermo_constraints}), and consequently, no contribution to the swimming speeds of the squirmers we consider. This is in contradiction to the results obtained through the weak $De$ expansion of a Giesekus fluid to $\mathcal{O} \left( De\right)$ where pushers and pullers swim faster and slower, respectively, than in a Newtonian fluid. This was discussed in the exchange between \citet{christov16} and \citet{decorato16b} described previously. 

To observe the effects of a fluid of grade three, we choose $\mathcal{P} = 3/2 $ (an arbitrary choice in as much as the physics of the problem is concerned). From \ref{thermo_constraints}, we know  $ -7\leq \mathcal{Q} \leq 5$.  Below, we plot the the swimming speeds for two cases: $\mathcal{P} = 3/2$, $\mathcal{Q} = -7$ and $\mathcal{P} = 3/2$, $\mathcal{Q} = 5$. 

\begin{figure}
\includegraphics[width =0.5 \textwidth]{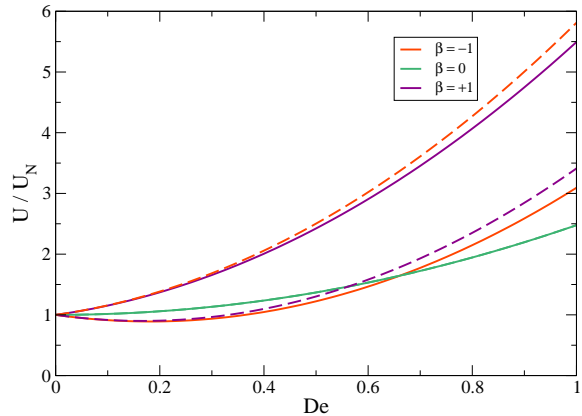}
\caption{Solid lines: $\mathcal{P} = 3/2$, $\mathcal{Q} = -7$. Dashed lines: $\mathcal{P} = 3/2$, $\mathcal{Q} = 5$. Solid and dashed lines for $\beta = 0$ overlap. Depending on the values of  $\mathcal{P}$, $\mathcal{Q}$, either of puller or pusher can swim faster or slower at small $De$. }
\label{third_swimming_comp_fig}
\end{figure}

From figure \ref{third_swimming_comp_fig} and equation \ref{third_swimming}, we see that depending on the value of $\mathcal{Q}$, either the puller or the pusher can swim faster than in a Newtonian fluid at $\mathcal{O} \left( De\right)$. The higher order correction, $\mathcal{O} \left(De^2 \right)$,  gives a positive contribution to the swimming speed.

In contrast to the results from the Giesekus fluid, the parameters in a fluid of grade three allow for a wider range of possibilities---either of puller or pusher can swim faster or slower at small values of the Deborah number.  About this range of possibilities, perhaps it is useful to recall the observation from  \citet{truesdell} that ``it is possible that two fluids of grade 3 could behave just alike in every viscometric test yet react altogether differently to some test of a different kind".    At higher $De$, all squirmers swim faster in fluids of grade three than in a Newtonian fluid, when in Giesekus fluids they would swim slower. 

\section{Conclusion}
We calculated the higher order corrections to swimming speeds in two viscoelastic fluids\textemdash Giesekus fluid and fluid of grade three. The higher order corrections significantly change the results from $\mathcal{O} \left( De\right)$ even at small values of $De$. We also see that the two fluids give qualitatively different results. Within the fluid of grade three, the presence of more free parameters allow for qualitatively different variations of the swimming speeds of the squirmers.  Clearly, the answer to what model to use for viscoelasticity depends on what we wish to model\textemdash in this, we are guided by experiments. 

\section*{Acknowledgement}
The authors thank Professor G.M. ``Bud" Homsy for a remark on our previous work. The remark became one of the seeds for the present work.

\bibliography{thesis}

\end{document}